\documentclass[prl,12pt]{revtex4}
\pdfoutput=1 
\usepackage{amsmath}
\usepackage{graphicx}
\usepackage{subfigure}
\usepackage{bm}
\usepackage{slashed}

\def\d{\mathrm{d}}

\begin{document}
\title{Predicting $\bar B^\circ\to \rho^\circ \gamma$ and $\bar B^\circ_s \to \rho^\circ \gamma$ using holographic AdS/QCD Distribution Amplitudes for the $\rho$ meson}

\author{M. Ahmady}
\affiliation{Department of Physics, Mount Allison University, Sackville, N-B. E46 1E6, Canada}
\email{mahmady@mta.ca} 
\author{R. Sandapen}
\affiliation{D\'epartement de Physique et d'Astronomie, Universit\'e de Moncton,
Moncton, N-B. E1A 3E9, Canada
\& \\
Department of Physics, Mount Allison University, Sackville, N-B. E46 1E6, Canada }
\email{ruben.sandapen@umoncton.ca}


\begin{abstract}
We derive holographic AdS/QCD Distribution Amplitudes for the transversely polarised $\rho$ meson and we use them to predict the branching ratio for the decays $\bar B^\circ\rightarrow \rho^\circ \gamma$ and $\bar B^\circ_s\rightarrow \rho^\circ \gamma$ beyond leading power accuracy in the heavy quark limit.  For $\bar B^\circ\rightarrow \rho^\circ \gamma$, our predictions agree with those generated using Sum Rules (SR) Distribution Amplitudes and with the data from the BaBar and Belle collaborations. In computing the weak annihilation amplitude which is power-supressed in $\bar B^\circ\rightarrow \rho^\circ \gamma$ but is the leading contribution in $\bar{B}^{\circ}_s\rightarrow \rho^\circ \gamma$, we find that, in its present form, the AdS/QCD DA avoids the end-point divergences encountered with the SR DA .  
\end{abstract}


\maketitle

\section{Introduction}
Flavor Changing Neutral Currents (FCNC) are excellent probes to the Standard Model and beyond. In particular, the $b\to(s,d) \gamma$ transitions are most important for the extraction of  the Cabibbo-Kobayashi-Maskawa (CKM) matrix elements as well as for the search of New Physics (NP) signals. The experimental measurements for the $b\to d \gamma$ transistion are currently available  for exclusive radiative $B$ decay to a $\rho$ meson, i.e. $B \to \rho \gamma$.  For a recent review of radiative $B$ decays, we refer to \cite{Hurth:2010tk}. 

 The theory of exclusive decays is complicated by their sensitivity to non-perturbative physics. The standard\footnote{Alternative frameworks can be found in reference \cite{Lu:2005yz,Ali:2001ez}.} theoretical framework is QCD factorization (QCDF)\cite{Bosch:2001gv,Beneke:2001at}  which states that,  to leading power accuracy in the heavy quark limit, the decay amplitude factorizes into perturbatively computable kernels and non-perturbative objects namely the $B\to V$ transition form factor, the meson couplings  and the leading twist Distribution Amplitudes (DAs) of the mesons. Traditionally the DAs for the vector meson are obtained from QCD Sum Rules\cite{Ball:1996tb,Ball:1998fj,Ball:1998ff}. The numerical values of the transition form factor and the tensor coupling of the vector meson are  obtained from light-cone Sum Rules or lattice QCD. The predictive power of QCDF is limited by the uncertainties associated with these non-perturbative quantities and also by power corrections to the leading  amplitude \cite{Antonelli:2009ws}. The computation of the power corrections is often problematic  due to the appearance of end-point divergences in convolution integrals that contribute to the decay amplitude \cite{Pecjak:2008gv,Antonelli:2009ws}.

Our goal in this paper is to use new holographic AdS/QCD DAs for the $\rho$ meson to compute the branching ratios for two exclusive radiative decays, namely $\bar B^\circ\rightarrow \rho^\circ \gamma$ and $\bar B^\circ_s\rightarrow \rho^\circ \gamma$, beyond leading power accuracy.  We derive the AdS/QCD DAs using a holographic AdS/QCD light-front  wavefunction for the $\rho$ meson \cite{deTeramond:2008ht}  which was recently shown to generate predictions for the cross-sections in diffractive $\rho$ meson production that are in agreement with the data collected at the HERA electron-proton collider \cite{Forshaw:2012im}. Reference \cite{Forshaw:2012im} also shows that the second moment of the AdS/QCD twist-$2$ DA  of the longitudinally polarised $\rho$ meson is in agreement with both Sum  Rules and lattice predictions.  Here we shall extend the comparison between AdS/QCD and Sum Rules for the twist-$2$ and twist-$3$ DAs  of the transversely polarized $\rho$ meson since both of these non-perturbative quantities are required to compute the  decay amplitudes for  $\bar B^\circ\rightarrow \rho^\circ \gamma$ and $\bar B^\circ_s\rightarrow \rho^\circ \gamma$ beyond leading power accuracy.

Theoretical predictions for the branching ratio of $\bar B^\circ\rightarrow \rho^\circ \gamma$, using the leading twist-$2$ SR DA for the $\rho$ meson,  can be found in references \cite{Bosch:2001gv,Ball:2006eu}. In both references, the leading power correction ($\mathcal{O} (\Lambda_{\mbox{\tiny{QCD}}}/m_b)$) due to annihilation is taken into account within QCDF  and the subleading  ($\mathcal{O} (\Lambda_{\mbox{\tiny{QCD}}}/m_b)^2$)  annihilation contributions are neglected. On the other hand, in reference \cite{Ball:2006eu}, two other classes of power corrections due to long-distance photon emission and soft gluon emission are also taken into account. Here we shall investigate the numerical importance of three additional subleading annihilation contributions to $\bar B^\circ\rightarrow \rho^\circ \gamma$, two of which turn out to be sensitive to the higher twist-$3$ DA of the $\rho$ meson.  In fact, the rare decay $\bar B^\circ_s\rightarrow \rho^\circ \gamma$ proceeds mainly via these four annihilation processes and  a theoretical prediction for its branching ratio using the SR twist-$3$ DA is available in reference \cite{Ahmady:2007ka}. However this prediction suffers from a large uncertainty because of end-point divergences encountered when computing those annihilation contributions sensitive to  the twist-$3$ DA for the $\rho$ meson.  We shall update this prediction using the twist-$3$ AdS/QCD DA which, as we shall see, avoids the end-point divergence problem. 

On the experimental side, the branching ratio for $\bar B^\circ\rightarrow \rho^\circ \gamma$ has been measured with increasing precision by the BaBar and Belle collaborations \cite{Taniguchi:2008ty}. On the other hand, the rare decay $\bar B^\circ_s\rightarrow \rho^\circ \gamma$ has not been measured experimentally but it is an interesting process to investigate at LHCb because of its sensitivity  to NP especially those which allow FCNC at tree level \cite{Ahmady:2007ka}.

We start by an outline of the computation of the amplitudes for both decays following references \cite{Bosch:2001gv,Ahmady:2007ka}, devoting attention to the quantities which are dependent on the DAs of the $\rho$ meson.

\section{Decay amplitudes}
The effective weak Hamiltonian for the underlying $b\rightarrow d\gamma$ transition  is given by 
\begin{equation}
{\cal{H}_{\rm eff}}=\frac{G_F}{\sqrt{2}}\sum_{p=u,c}V_{pq}^*V_{pb}[ C_1Q_1^p+C_2Q_2^p+\sum_{i=3}^{8} C_iQ_i]
\label{WH}
\end{equation}
where $q=d$ for  $\bar B^\circ\rightarrow \rho^\circ \gamma$ and $q=s$ for $\bar B^\circ_s\rightarrow \rho^\circ \gamma$. $Q_1^p$ and $Q_2^p$ are the current-current operators, $Q_{i=3..6}$ are the QCD penguin operators and $Q_7$ and $Q_8$ are the electromagnetic and chromomagnetic operators.  The coeffecients $C_i$ are  the  perturbatively known Wilson coeffecients and $V_{ij}$ are the CKM matrix elements. In this paper, we shall use the NLO Wilson coeffecients given in \cite{Ball:2006eu,Buchalla:1995vs}\footnote{Note that $C_{1,2}$(here)$=C_{2,1}$(reference \cite{Ball:2006eu}).} and we use the numerical values of  CKM matrix elements given in reference \cite{Beringer:1900zz}. 

We start with  the amplitude for the decay $\bar B^\circ\rightarrow \rho^\circ \gamma$. At leading power accuracy in the heavy quark limit and to all orders in the strong coupling $\alpha_s$,  the matrix elements of these operators factorize as\cite{Bosch:2001gv}:
\begin{equation}
\langle \rho(P,e_T) \gamma (q,\epsilon ) | Q_i | \bar{B} \rangle = [ F^{B\rightarrow \rho} T_i^I  + \int_0^1 \d \zeta\; \d z \; \Phi_B(\zeta) T_i^{II}(\zeta,z)  \phi^{\perp}_{\rho} (z)] \cdot \epsilon 
\label{factorization}
\end{equation}
i.e. into perturbatively computable  hard-scattering kernels $T_i^I$  and $T_i^{II}$ and three non-perturbative quantities namely the transition form factor $F^{B\rightarrow \rho}$, the leading twist DA of the $B$ meson, $\Phi_B(\zeta) $, and the twist-$2$ DA of $\rho$ meson, $\phi_{\rho}^{\perp} (z)$. In equation \eqref{factorization}, $P$ and $e_T $ are the $4$-momentum and polarization vector of the $\rho$ meson while $q$ and $\epsilon$ are the $4$-momentum and polarization vector of the photon. The form factor $F^{B\rightarrow \rho}$ is obtained from light-cone QCD Sum Rules \cite{Ball:2006eu}. The second term in equation \eqref{factorization} describes mechanisms involving the spectator quark, hence its dependence on the DAs of the mesons. In what follows, we shall only need the first inverse moment of  $\Phi_B(\zeta) $ which is parametrized as $M_B/\lambda_B$ where $\lambda_B = \mathcal{O} (\Lambda_{\mbox{\tiny{QCD}}})$ \cite{Bosch:2001gv}.

At zeroth order in $\alpha_s$, the leading power amplitude  is given by 
\begin{equation}
{\cal A}_{\bar B^\circ\to\rho^\circ\gamma}^{\tiny{\mbox{Leading}}}= \frac{G_F}{\sqrt{2}}  \sum_{p=u,c}V_{pd}^*V_{pb} C_7 \langle Q_7 \rangle 
\label{LOA}
\end{equation}
where $\langle Q_7 \rangle$ is the matrix element of the operator $Q_7$, i.e.
\begin{equation}
\langle \rho (P,e) \gamma(q,\epsilon) | Q_7 | \bar{B}^{\circ} \rangle = -\frac{em_b(\mu) F^{B\rightarrow \rho} } {2\sqrt{2} \pi^2} \left [\varepsilon_{\mu\nu\alpha\beta}\epsilon^\mu e_T^\nu P^\alpha
q^\beta + i((\epsilon \cdot e_T) (P \cdot q)-(\epsilon \cdot P) (q \cdot e_T))\right] 
\label{Q7}
\end{equation}
where $e=\sqrt{4 \pi \alpha_{\rm{em}}}$ and $m_b(\mu)$ is the running mass of the $b$ quark evaluated at the hard scale $\mu=m_b$. 

At order $\alpha_s$, the leading power amplitude becomes\cite{Bosch:2001gv}  
\begin{equation}
{\cal A}_{\bar B^\circ\to\rho^\circ\gamma}^{\tiny{\mbox{Leading}}}= \frac{G_F}{\sqrt{2}} \sum_{p=u,c}V_{pd}^*V_{pb} a_7^p \langle Q_7 \rangle 
\label{NLOA}
\end{equation}
with
\begin{eqnarray}
a_7^p &=&C_7+\frac{\alpha_s(\mu )C_F}{4\pi} [C_1(\mu)G_1(s_p)+C_8(\mu )G_8] \nonumber \\
&+&\frac{\alpha_s(\mu_{h})C_F}{4\pi}[C_1(\mu_{h} )H_1(s_p,\mu_{h})+C_8(\mu_{h} )H_8(\mu_{h})] \;.
\label{a7}
\end{eqnarray}
In equation \eqref{a7}, the strong coupling is evaluated at two different scales: $\mu=m_b$ and a hadronic scale $\mu_{h}=\sqrt{\Lambda_{\tiny{\mbox{QCD}}} \mu}$. The functions $G_1$ and $H_1$ depend on $s_p=(m_p/m_b)^2$ where $m_p$ is the quark mass in loops contributing at next-to-leading order accuracy in $\alpha_s$. The hard scattering functions $G_1(s_p)$ and $G_8$  are given explicitly in reference \cite{Bosch:2001gv}. Here, we focus on the functions $H_8(\mu_h)$ and  $H_1(s_p,\mu_h)$ which depend on the twist-$2$ DA of the $\rho$ meson. The function $H_1(s_p,\mu)$ is given by \cite{Bosch:2001gv}
\begin{equation}
H_1(s_p,\mu)=-\left(\frac{2 \pi^2 f_B f_{\rho}^{\perp}(\mu) }{3 N_c M_B^2} \right) \left(\frac{M_{B}}{\lambda_B} \right) I_1^{\mbox{\tiny{tw2}}}(s_p,\mu)
\label{H1sf}
\end{equation}
where $f_B$ is the decay constant of the $B$ meson which is obtained from lattice QCD \cite{Na:2012kp,Bazavov:2011aa}. 
$I_1^{\mbox{\tiny{tw2}}}(s_p,\mu)$ is a convolution of the twist-$2$ DA with a hard scattering kernel, i.e.
\begin{equation}
I_1^{\mbox{\tiny{tw2}}}(s_p,\mu)=\int_0^1 \d z \; h(s_p,\bar{z}) \phi_{\rho}^{\perp} (z,\mu) 
\label{I1}
\end{equation}
where the hard scattering kernel is given by
\begin{equation}
h(s_p, \bar{z})=\left \{ \frac{4s_p}{\bar{z}^2} \left [L_2\left (\frac{2}{1-\sqrt{\frac{\bar{z}-4s_p+i\epsilon}{\bar{z}}}}\right )+
L_2\left (\frac{2}{1+\sqrt{\frac{\bar{z}-4s_p+i\epsilon}{\bar{z}}}}\right )\right ]-\frac{2}{\bar{z}} \right \}
\end{equation}
with $L_2$ being the dilogarithmic function and $\bar{z}=1-z$. The function $H_8(\mu)$ is given by  
\begin{equation}
H_8(\mu) = \left(\frac{4 \pi^2 f_B f_{\rho}^{\perp}(\mu)}{3 N_c F^{B \to \rho } M_B^2} \right) \left(\frac{M_{B}}{\lambda_B} \right) I_2^{\mbox{\tiny{tw2}}}(\mu)
\label{H8}
\end{equation}
where $I_2^{\mbox{\tiny{tw2}}}(\mu)$ is the first inverse moment of the twist-$2$ DA, i.e.
\begin{equation}
I_2^{\mbox{\tiny{tw2}}}(\mu)=\int_0^1 \d z \; \frac{\phi_{\rho}^{\perp}(z,\mu)} {z} \;.
\label{I2}
\end{equation}
We note that if $m_b \gg m_p$, then 
\begin{equation}
I_1^{\mbox{\tiny{tw2}}}(s_p,\mu) \approx I^{\mbox{\tiny{tw2}}}_1(0,\mu) =-2 I_2^{\mbox{\tiny{tw2}}}(\mu) 
\end{equation}
so that both $H_1$ and $H_8$ become simply proportional to the first inverse moment of the twist-$2$ DA of the $\rho$ meson. In practice, this approximation is not justified for a charm loop, i.e. when $p=c$ and we do not make it here. In what follows, we shall take $m_{u,d}=0.14$ GeV \cite{Forshaw:2012im}, $m_c=1.3$ GeV and $m_b=4.2$ GeV. 

\begin{figure}
\centering
\includegraphics[width=.50\textwidth]{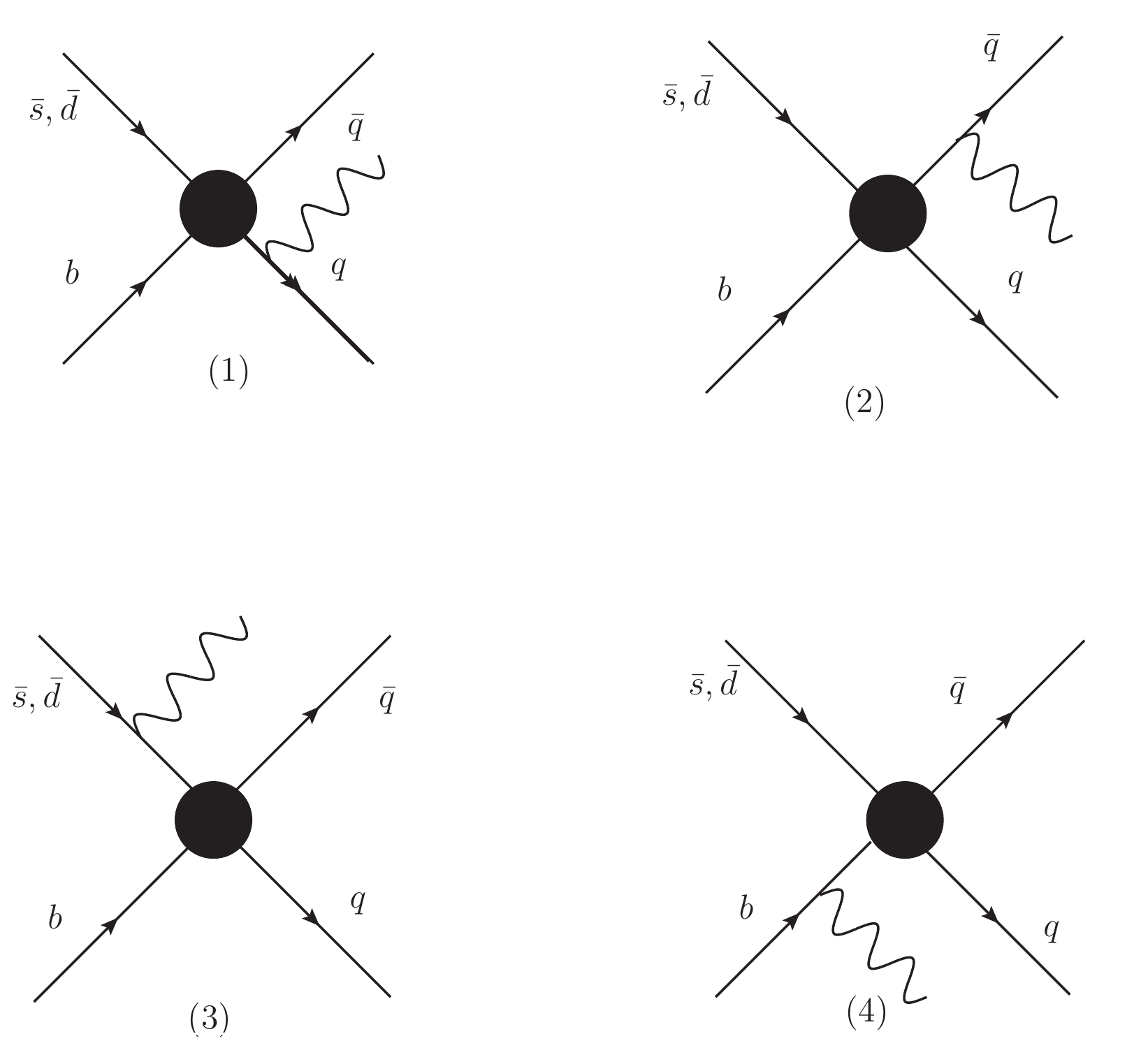}
\caption{Annihilation processes contributing to the decay $\bar{B}^\circ \to \rho^\circ \gamma$ and $\bar B^\circ_s \to\rho^\circ\gamma$. The diagrams (1),(2) and (4) are power-suppressed compared to the diagram (3).  The contributions (1) and (2) are sensitive to the twist-$3$ DA of the $\rho$ meson. }
\label{fig:annihilation}
\end{figure}

In $\bar B^\circ\rightarrow \rho^\circ \gamma$, all annihilation topologies are suppressed by at least one power of $\Lambda_{\mbox{\tiny{QCD}}}/m_b$ \cite{Bosch:2001gv}. The leading annihilation contribution is given by \cite{Bosch:2001gv}
\begin{equation}
{\cal A}_{\tiny{\mbox{annihilation}}}^{\tiny{\mbox{leading}}} (\bar B^\circ\rightarrow \rho^\circ \gamma)= \frac{G_F}{\sqrt{2}} V_{ud}^* V_{ub} \left(C_2 + \frac{1}{N_c}C_1 \right) b_d \langle Q_7 \rangle
\label{leading-ann-BB}
\end{equation}
with 
\begin{equation}
b_d= \frac{2 \pi f_B f_{\rho} M_{\rho}}{F^{B \to \rho} M_B^2 \lambda_B} \;.
\label{bd}
\end{equation}
This leading contribution can be taken into account  by adding an extra term to the coeffecient $a_7^u$ in the leading power amplitude given by equation \eqref{NLOA}:
\begin{equation}
a_7^u \rightarrow a_7^u + b_d \left( C_2 + \frac{1}{N_c}C_1 \right) 
\end{equation}
where $b_d$ is given by equation \eqref{bd}. The leading annihilation contribution corresponds to the annihilation diagram in which the photon is radiated off the spectator quark of the $B$ meson, i.e. the third diagram in figure \ref{fig:annihilation}. 
Here we wish to investigate the numerical importance of  the three other subleading annihilation contributions shown in figure \ref{fig:annihilation}. In fact, the four annihilation diagrams of figure \ref{fig:annihilation} are the dominant contributions to the decay $\bar B^\circ_s\rightarrow \rho^\circ \gamma$
\cite{Ahmady:2007ka}.  The total annihilation amplitude is given by\cite{Ahmady:2007ka} 
\begin{equation}
{\cal A}_{d(s)}^{\mbox{\tiny{annihilation}}}= \frac{e G_F}{\sqrt{2}} V^*_{td(s)}V_{tb}f_{B_{(s)}} f_\rho M_\rho ({\cal A}_{d(s)}^{1} + {\cal A}_{d(s)}^{2} + {\cal A}_{d(s)}^{3} + {\cal A}_{d(s)}^{4} )
\label{Anni-tot}
\end{equation}
where to zeroth order in $\alpha_s$,
\begin{equation}
{\cal A}_{d(s)}^{1}  + {\cal A}_{d(s)}^{2}=  2C_{12} [I_{1(s)}^{\mbox{\tiny{tw3}}} (\mu) -I_{2(s)}^{\mbox{\tiny{tw3}}} (\mu)]\varepsilon_{\mu\nu\alpha\beta} \epsilon^\mu e_T^\nu P^\alpha q^\beta
\label{A1}
\end{equation}
with 
\begin{equation}
 I_{1(s)}^{\mbox{\tiny{tw3}}} (\mu) = \int_0^1 \d z \; \frac{g_\rho^{\bot (v)}(z,\mu)}{zM_{B_{(s)}}^2+z\bar zM_\rho^2-m_f^2}
\label{I3}
\end{equation}
and
\begin{equation}
I_{2(s)}^{\mbox{\tiny{tw3}}} (\mu) =\int_0^1 \d z \; \frac{zg_\rho^{\bot (v)}(z,\mu)}{zM_{B_{(s)}}^2+z\bar
  zM_\rho^2-m_f^2} 
\label{I4}
\end{equation}
while
\begin{eqnarray}
{\cal A}_{d(s)}^{3} =  C_{34} \left( \frac{1}{2 E_{\gamma} \lambda_{B_{(s)}}} \right)   \left\{\varepsilon_{\mu\nu\alpha\beta} \epsilon^\mu e_T^\nu P^\alpha q^\beta + i [(\epsilon \cdot e_T) (P \cdot q)-(\epsilon \cdot P) (q \cdot e_T)] \right. \\ \nonumber 
\left. + i \frac{\lambda_{B_{(s)}}}{M_{B_{(s)}}} \left[2 (\epsilon \cdot P) (q \cdot e_{T}) - M_{B_{(s)}}^2 \left(1 + \frac{m_{d(s)}}{M_{B_{(s)}}}\right)(\epsilon \cdot e_T)\right] \right\}
\label{A3}
\end{eqnarray}
and
\begin{eqnarray}
{\cal A}_{d(s)}^{4} = C_{34} \left( \frac{1}{2 M_{B_{(s)}} E_{\gamma}} \right)  \left \{ \varepsilon_{\mu\nu\alpha\beta} \epsilon^\mu e_T^\nu P^\alpha q^\beta - i [(\epsilon \cdot e_T) (P \cdot q)-(\epsilon \cdot P) (q \cdot e_T)] \right. \\ \nonumber
\left. - i \left[2 (\epsilon \cdot P) (q \cdot e_{T}) - M_{B_{(s)}}^2 \left(1 + \frac{m_{b}}{M_{B_{(s)}}}\right)(\epsilon \cdot e_T)\right] \right\}
\label{A4}
\end{eqnarray}

In the above equations, $C_{12}$ and $C_{34}$ are the combinations of the Wilson coefficients\footnote{For notational simplicity, we suppress the dependence of the Wilson coeffecients on the scale $\mu=m_b$.}\cite{Ahmady:2007ka}:  
\begin{eqnarray}
  C_{12} &=& \frac{1}{\sqrt{2}}\left [ \frac{2}{3}\left (C_2+\frac{C_1}{3}\right )\frac{V_{ub}V^*_{ud(s)}}{V_{tb}V^*_{td(s)}}+\left (C_3+\frac{C_4}{3}-C_5-\frac{C_6}{3}\right )\right ],\\
  C_{34} &=&  \frac{1}{\sqrt{2}}\left [ \frac{1}{3}\left (C_2+\frac{C_1}{3}\right )\frac{V_{ub}V^*_{ud(s)}}{V_{tb}V^*_{td(s)}}\right ] 
\label{C1C2C3C4}
\end{eqnarray}
and $E_{\gamma(s)}$ is the energy of the photon in the $B_{(s)}$ meson rest frame, i.e.
\begin{equation}
E_{\gamma(s)}=\frac{M_{B_{(s)}}}{2} \left(1-\frac{M_{\rho}^2}{M_{B_{(s)}}^2} \right) \;.
\label{Egamma}
\end{equation}
 The quantity $\lambda_{B_{s}}$ is analogous to $\lambda_B$, i.e. it parametrizes the first inverse moment of the $\bar B^\circ_s$ meson  DA. As expected the amplitudes ${\cal A}_{d(s)}^{1,2}$, corresponding to annihilation topologies in which the photon is radiated off the light quark or antiquark of the $\rho$, are sensitive to the twist-$3$ DA of the $\rho$ meson. 
Note that the annihilation amplitude ${\cal A}_{d}^3$, evaluated  to leading power in the heavy quark limit, coincides with equation \eqref{leading-ann-BB} which is the leading annihilation contribution given in reference \cite{Bosch:2001gv}.

The total decay amplitude for $\bar B^\circ\to\rho^\circ\gamma$ is then
\begin{equation}
{\cal A} (\bar B^\circ\to\rho^\circ\gamma) = {\cal A}_{\bar B^\circ\to\rho^\circ\gamma}^{\tiny{\mbox{Leading}}} +  {\cal A}_{d}^{\tiny{\mbox{Annihilation}}} 
\label{AtotB0}
\end{equation}
where ${\cal A}_{\bar B^\circ\to\rho^\circ\gamma}^{\tiny{\mbox{Leading}}}$ and ${\cal A}_{d}^{\tiny{\mbox{Annihilation}}} $ are given by equation \eqref{NLOA} and \eqref{Anni-tot} respectively. 
On the other hand, the total decay amplitude for $\bar{B}^\circ_{s} \to \rho^\circ \gamma$ is given by
\begin{equation}
{\cal A} (\bar B^\circ_s\to\rho^\circ\gamma) =   {\cal A}_{s}^{\tiny{\mbox{Annihilation}}} 
\label{AtotBs}
\end{equation}
where $ {\cal A}_{s}^{\tiny{\mbox{Annihilation}}} $ is given by equation \eqref{Anni-tot}. 
 
In order to compute the decay amplitudes given by equations \eqref{AtotB0} and \eqref{AtotBs}, we must specify the  twist-$2$ DA $\phi_{\rho}^{\perp}$  in equations \eqref{I1} and \eqref{I2} as well as  the  twist-$3$ DA $g_{\rho}^{\perp (v)}$ appearing in equations \eqref{I3} and \eqref{I4}.  We also need to specify the numerical value of the tensor coupling $f_{\rho}^{\perp}$ which appears in equations \eqref{H1sf} and \eqref{H8}.  We shall do this in the next section using a holographic AdS/QCD light-front wavefunction for the $\rho$ meson. The numerical values of the decay constants $f_{B_{(s)}}$, the parameters $\lambda_{(s)}$ and the form factor $F_{B \to \rho}$ are shown  in table \ref{tab:params}. 

\begin{table}[h]
\begin{center}
\[
\begin{array}
[c]{|c|c|}
\hline
 \mbox{Parameter} &\mbox{Numerical value} \\ \hline
\lambda_{B_{(s)}} & 0.51 \pm 0.12~ (0.60 \pm 0.20) ~\mbox{GeV}\\ \hline
f_{B_{(s)}} & 190.6 \pm 47 ~(227.6 \pm 5.0) ~\mbox{MeV}\\ \hline
F^{B\to \rho}& 0.27 \pm 0.04 \\ \hline
\end{array}
\]
\end{center}
\caption {The non-perturbative input parameters obtained from QCD Sum Rules or lattice QCD \cite{Ball:2006eu,Bazavov:2011aa,McNeile:2011ng,Na:2012kp,Laiho:2009eu}. We shall take the central values of these parameters to compute our predictions for the branching ratios.}
\label{tab:params}
\end{table}

\section{Holographic AdS/QCD DAs and couplings of the $\rho$ meson}

Distribution Amplitudes parameterize  the operator product expansion of vacuum-to-meson
transition matrix
elements of quark-antiquark non-local gauge invariant operators at light-like
separations.  At equal light-front time $x^+=0$ and in the light-front gauge $A^+=0$, we have \cite{Ball:1996tb,Ball:1998fj}
\begin{eqnarray}
\langle 0|\bar q(0)  \gamma^\mu q(x^-)|\rho
(P,\lambda)\rangle
&=& f_\rho M_\rho
\frac{e_{\lambda} \cdot x}{P^+x^-}\, P^\mu \int_0^1 \d u \; e^{-iu P^+x^-}
\phi_{\rho}^\parallel(u,\mu)
\nonumber \\
&\hspace{-1.0cm}+&\hspace{-0.5cm} f_\rho M_\rho
\left(e_{\lambda}^\mu-P^\mu\frac{e_{\lambda} \cdot
x}{P^+x^-}\right)
\int_0^1 \d u \; e^{-iu P^+x^-} g^{\perp (v)}_{\rho} (u,\mu)  \;,
\label{DA:phiparallel-gvperp}
\end{eqnarray}
\begin{equation}
\langle 0|\bar q(0) [\gamma^\mu,\gamma^\nu] q(x^-)|\rho (P,\lambda)\rangle =2 f_\rho^{\perp} (e^{\mu}_{\lambda} P^{\nu} - e^{\nu}_{\lambda} P^{\mu}) \int_0^1 \d u \; e^{-iuP^+ x^-} \phi_{\rho}^{\perp} (u, \mu)
\label{DA:phiperp}
\end{equation}
and
\begin{equation}
\langle 0|\bar q(0) \gamma^\mu \gamma^5 q(x^-)|\rho (P,\lambda)\rangle =-\frac{1}{4} \epsilon^{\mu}_{\nu\rho\sigma} e_{\lambda}^{\nu} P^{\rho} x^{\sigma}  f_\rho M_{\rho} \int_0^1 \d u \; e^{-iuP^+ x^-} g^{\perp (a)}_{\rho}(u, \mu)
\label{DA:gaperp}
\end{equation}
for the vector, tensor and axial-vector current respectively. The polarization vectors $e_{\lambda}$ are chosen as
\begin{equation}
e_{L}=\left(\frac{P^{+}}{M_{\rho}},-\frac{M_{\rho}}{P^{+}},0_{\perp} \right) ~\hspace{1cm}~\mbox{and}\hspace{1cm}  e_{T(\pm)}=\frac{1}{\sqrt{2}}\left(0,0, 1, \pm i \right)
\end{equation}
where $P^+$ is the ``plus'' component of the $4$-momentum of the meson given by
\begin{equation}
P^{\mu}=\left(P^{+},\frac{M_{\rho}^2}{P^{+}} ,0_{\perp} \right) \;.
\end{equation}
All four DAs satisfy the normalization condition
\begin{equation}
 \int_0^1 \d z \; \varphi  (z,\mu) = 1
\label{NormDA}
\end{equation}
where $\varphi=\{\phi_{\rho}^{\parallel,\perp},g^{\perp (v,a)}_{\rho} \}$ so that for a vanishing light-front distance $x^-= 0$, the definitions of the vector coupling $f_{\rho}$ and tensor coupling $f^{\perp}_{\rho}$ are recovered, i.e.
\begin{equation}
\langle 0|\bar q(0)  \gamma^\mu q(0)|\rho
(P,\lambda)\rangle =f_\rho M_\rho e_\lambda^{\mu}
\label{frho}
\end{equation}
and
\begin{equation}
\langle 0|\bar q(0) [\gamma^\mu,\gamma^\nu] q(0)|\rho (P,\lambda)\rangle =2 f_\rho^{\perp}  (e^{\mu}_{\lambda} P^{\nu} - e^{\nu}_{\lambda} P^{\mu}) \;.
\label{fperp}
\end{equation}
The vector coupling $f_{\rho}$ is accessible experimentally via the leptonic decay width of the $\rho$ meson \cite{Ball:2006eu} 
\begin{equation}
f_\rho  = \left(  \frac{6\Gamma_{e^{+}e^{-}} M_\rho}{4 \pi
    \alpha_{\mathrm{em}}^2} \right)^{1/2}
\end{equation}
 where $\Gamma_{e^{+}e^{-}}=7.04 \pm 0.06~\mathrm{keV}$ \cite{Beringer:1900zz}.   On the other hand, the tensor coupling  $f_{\rho}^{\perp}$ is not measured experimentally but is predicted theoretically by QCD Sum Rules and lattice QCD.

It follows from equations \eqref{DA:phiparallel-gvperp}, \eqref{DA:phiperp} and \eqref{DA:gaperp} that the twist-$2$ DAs are given by
\begin{equation}
f_{\rho} \phi_{\rho}^{\parallel}(z,\mu)= \int \d x^- \; e^{i zP^+
x^-}\langle 0|\bar q(0)
\gamma^+ q(x^-)|\rho
(P,L) \rangle 
\label{phiparallel}
\end{equation}
and
\begin{equation}
f_{\rho}^{\perp} \phi_{\rho}^{\perp}(z,\mu)= \frac{1} {2} \int \d x^- \; e^{i zP^+
x^-}
\langle
0|\bar q(0)
[e^{*}_{T(\pm)}.\gamma,\gamma^+] q(x^-)|\rho
(P,T(\pm)) \rangle
\label{phiperp}
\end{equation}
 while the twist-$3$ DAs are given by
\begin{equation}
f_{\rho} g_{\rho}^{\perp (v)}(z,\mu)= \frac{P^+}{M_{\rho}}\int \d x^-  \; e^{i zP^+
x^-}
\langle
0|\bar q(0)
e^{*}_{T(\pm)}.\gamma q(x^-)|\rho
(P,T(\pm)) \rangle 
\label{gvperp}
\end{equation}
and
\begin{equation}
f_{\rho} \frac{\d g_{\rho}^{\perp (a)}}{\d z}(z,\mu)= \mp \frac{2 P^+}{M_{\rho}}\int \d x^- \;  e^{i zP^+
x^-}
\langle
0|\bar q(0)
e^{*}_{T(\pm)}.\gamma \gamma^5 q(x^-)|\rho
(P,T(\pm)) \rangle \;.
\label{gaperp}
\end{equation}

To relate the DAs  to the light-front wavefunctions of the $\rho$ meson, we use the relation \cite{Forshaw:2011yj}
\begin{eqnarray}
\label{general-relation}
P^+\int \d x^- e^{ix^-zP^+} \langle 0 | \bar{q}(0)  \Gamma
q(x^-)
|\rho(P,\lambda) \rangle &=& \frac {N_c}{4\pi} \sum_{h,\bar{h}}
\int^{|\mathbf{k}| < \mu}
\frac{\d^2\mathbf{k}}{(2\pi)^2} \;
S^{\rho,\lambda}_{h,\bar{h}}(z,\mathbf{k}) \phi_{\lambda} (z,\mathbf{k}) \nonumber \\
&\times&
\left \{ \frac{\bar{v}_{\bar{h}}((1-z)P^{+},-\mathbf{k})}{\sqrt{(1-z)}}
\Gamma \frac{u_h(zP^+,\mathbf{k})}{\sqrt{z}} \right \}
\end{eqnarray}
where we have identified the renormalization scale $\mu$ as a cut-off on the transverse momentum of the quark \cite{Forshaw:2011yj} and $\phi_{\lambda}(z,\mathbf{k})$ is the meson light-front wavefunction in momentum space. A two-dimensional Fourier transform of $\phi_{\lambda}(z,\mathbf{k})$ gives the light-front wavefunction, $\phi_{\lambda}(z,\mathbf{r})$, in configuration space. The light-front wavefunctions can be modelled \cite{Kulzinger:1998hw,Forshaw:2003ki,Nemchik:1996cw} or extracted from data \cite{Forshaw:2011yj}. Here we use the AdS/QCD holographic wavefunction predicted in \cite{Brodsky:2007hb,Brodsky:2008pg} and which can be written as
\cite{Vega:2009zb}
\begin{equation}
\phi_{\lambda} (z,\zeta)= \mathcal{N}_{\lambda} \frac{\kappa}{\sqrt{\pi}}\sqrt{z(1-z)} \exp \left(-\frac{\kappa^2 \zeta^2}{2}\right) \exp\left(-\frac{m_f^2}{2\kappa^2 z (1-z)} \right)
\label{AdS-QCD-wfn}
\end{equation}
where $\zeta=\sqrt{z(1-z)} r$ is the transverse distance between the quark and antiquark at equal light-front time\footnote{ The transverse separation between the quark and antiquark at equal ordinary time is $r$.}  and is the variable that maps onto the fifth dimension of AdS space \cite{deTeramond:2008ht,Brodsky:2012je,deTeramond:2012cs}. The AdS/QCD wavefunction given by  equation \eqref{AdS-QCD-wfn} is obtained using the  soft-wall model \cite{Karch:2006pv} to simulate confinement  and in that case the parameter $\kappa=M_{\rho}/\sqrt{2}$ where $M_\rho$ is the mass of the $\rho$ meson.  This AdS/QCD wavefunction has recently been used within the dipole model to generate parameter-free\footnote{The quark mass $m_f$ is chosen as $0.14$ GeV which is the value used in the dipole fits to the structure function $F_2$ data \cite{Soyez:2007kg,Forshaw:2006np,Forshaw:2004vv}. We shall also use this value here.} predictions for diffractive $\rho$ meson electroproduction that are in agreement with the HERA data \cite{Forshaw:2012im}. As discussed in reference \cite{Forshaw:2012im}, the normalization $\mathcal{N}_{\lambda}$ of the AdS/QCD wavefunction is allowed to depend on the polarisation of the meson $\lambda=L,T$.

Going back to  equation \eqref{general-relation}, the spinor wavefunctions $S_{h,\bar{h}}^{\rho,\lambda}(z,\mathbf{k})$ are given by \cite{Forshaw:2011yj}
\begin{equation}
S_{h,\bar{h}}^{\rho,L}(z,\mathbf{k})=
\left[M_{\rho} +  \frac{m_{f}^2 +  \mathbf{k}^{2}}{z(1-z)M^2_{\rho}} \right]
\delta_{h,-\bar{h}}
\label{SL}
\end{equation}
and
\begin{equation}
S_{h,\bar{h}}^{\rho,T(\pm)}(z,\mathbf{k})=
\frac{\sqrt{2}}{z(1-z)} \{ [(1-z) \delta_{h\mp,\bar{h}\pm} - z
\delta_{h\pm,\bar{h}\mp} ]k e^{\pm i \theta_k} \mp m_f
\delta_{h\pm,\bar{h}\pm} \} 
\label{ST}
\end{equation}
while $\Gamma$ stands for $\gamma^+$,  $[e^{*}_{T(\pm)}.\gamma,\gamma^+]$, $e^{*}_{T(\pm)}.\gamma$ or $e^{*}_{T(\pm)}.\gamma \gamma^5$. The matrix element in curly brackets of equation \eqref{general-relation} can then be evaluated explicitly for each case using the light-front spinors of reference \cite{Lepage:1980fj}:
\begin{equation}
\frac{\bar{v}_{\bar{h}}}{\sqrt{(1-z)}}\gamma^+\frac{u_h}{\sqrt{z}} =2 P^+ \delta_{h,-\bar{h}} \;,
\label{gammap}
\end{equation}
\begin{equation}
\frac{\bar{v}_{\bar{h}}}{\sqrt{(1-z)}}
 [e^{*}_{T(\pm)}.\gamma,\gamma^+] \frac{u_h}{\sqrt{z}} = \mp 4\sqrt{2} P^+ \delta_{h\pm,\bar{h} \pm} \;,
\label{commutator}
\end{equation}
\begin{equation}
\frac{\bar{v}_{\bar{h}}}{\sqrt{(1-z)}}
 e^{*}_{T(\pm)}.\gamma \frac{u_h}{\sqrt{z}} =
\frac{\sqrt{2}}{z(1-z)} \{ [(1-z) \delta_{h\mp,\bar{h}\pm} - z \delta_{h\pm,\bar{h}\mp} ] k e^{\mp i \theta_k} \mp m_f
\delta_{h\pm,\bar{h}\pm} \} 
\label{gammaperp}
\end{equation}
and
\begin{equation}
\frac{\bar{v}_{\bar{h}}}{\sqrt{(1-z)}}
 e^{*}_{T(\pm)}.\gamma \gamma^5 \frac{u_h}{\sqrt{z}} =
\frac{\sqrt{2}}{z(1-z)} \{ \mp [z \delta_{h\pm,\bar{h}\mp} + (1-z)
\delta_{h\mp,\bar{h}\pm}] k e^{\mp i \theta_k} + (1-2z) m_f \delta_{h\pm,\bar{h}\pm} \} \;.
\label{gammaperpgamma5}
\end{equation}
We are then able to deduce that
\begin{equation}
\phi_{\rho}^\parallel(z,\mu) =\frac{N_c}{\pi f_{\rho} M_{\rho}} \int \d
r \; \mu
J_1(\mu r) [M_{\rho}^2 z(1-z) + m_f^2 -\nabla_r^2] \frac{\phi_L(r,z)}{z(1-z)} \;,
\label{phiparallel-phiL}
\end{equation}
\begin{equation}
\phi_{\rho}^{\perp}(z,\mu) =\frac{N_c m_f}{\pi f_{\rho}^{\perp}} \int \d
r \; \mu
J_1(\mu r) \frac{\phi_T(r,z)}{z(1-z)} \;,
\label{phiperp-phiT}
\end{equation}
\begin{equation}
g_{\rho}^{\perp (v)}(z,\mu)=\frac{N_c}{2 \pi f_{\rho} M_{\rho}} \int \d r \; \mu
J_1(\mu r)
\left[ m_f^2 - (z^2+(1-z)^2) \nabla_r^2 \right] \frac{\phi_T(r,z)}{z^2 (1-z)^2
}
\label{gvperp-phiT}
\end{equation}
and
\begin{equation}
\frac{\d g_{\rho}^{\perp (a)}}{\d z}(z,\mu)=\frac{\sqrt{2} N_c}{\pi f_{\rho} M_{\rho}} \int \d r \; \mu
J_1(\mu r)
(1-2z) [m_f^2 - \nabla_r^2]\frac{\phi_T(r,z)}{z^2 (1-z)^2} \;.
\label{gaperp-phiT}
\end{equation}
Equations \eqref{phiparallel-phiL} and \eqref{gvperp-phiT} were derived in reference \cite{Forshaw:2011yj} where the light-front wavefunctions $\phi_{\lambda}(r,z)$ were extracted from data.  Equations  \eqref{phiperp-phiT} and \eqref{gaperp-phiT}  are new results which show how the twist-$2$ and twist-$3$ DAs of the transversely polarised $\rho$ meson are related to its light-front wavefunction.

We are also able to express the vector and tensor couplings $f_{\rho}$ and $f_{\rho}^{\perp}$ in terms of the light-front wavefunctions. From the definitions \eqref{frho} and \eqref{fperp}, it follows that
\begin{equation}
\langle 0|\bar q(0)  e_L^* \cdot \gamma q(0)|\rho
(P,L\rangle =f_\rho M_\rho 
\label{frho1}
\end{equation}
and
\begin{equation}
\langle 0|\bar q(0) [e_{T(\pm)}^* \cdot \gamma,\gamma^+] q(0)|\rho (P,T)\rangle =2 f_\rho^{\perp}  P^{+}  \;.
\label{frhoperp1}
\end{equation}
After expanding the left-hand-sides of equations \eqref{frho1} and \eqref{frhoperp1}, we obtain the decay width constraint \cite{Forshaw:2003ki}
\begin{equation}
f_\rho = \frac{N_c}{M_\rho \pi}  \int_0^1 \d z \;
\left.[z(1-z)M^{2}_{\rho} + m_{f}^2 -\nabla_{r}^{2}]
\frac{\phi_L(r,z)}{z(1-z)}
\right|_{r=0}
\label{vector-decay}
\end{equation}
 and
\begin{equation}
f_{\rho}^{\perp}(\mu) =\frac{m_f N_c}{\pi} \int_0^1 \d z \; \int \d r \; \mu J_1(\mu r)  \frac{\phi_T(r,z)}{z(1-z)} 
\label{tensor-decay-mu}
\end{equation}
respectively.
Note that equations \eqref{vector-decay} and \eqref{tensor-decay-mu} can also be obtained by inserting equations \eqref{phiparallel-phiL} and \eqref{phiperp-phiT} into  
the normalization conditions on the twist-$2$ DAs, i.e. into
\begin{equation}
\int_0^1 \d z \; \phi_{\rho}^{\parallel}(z,\infty) = 1 
\end{equation}
and 
\begin{equation}
\int_0^1 \d z \; \phi_{\rho}^{\perp}(z,\mu) = 1 
\end{equation}
respectively.

\section{Comparison to DAs and couplings from Sum Rules}
Inserting equation \eqref{AdS-QCD-wfn} in equations \eqref{vector-decay} and \eqref{tensor-decay-mu}, we can compute the AdS/QCD predictions for the vector and tensor couplings of the $\rho$ meson. We compare our predictions to experiment, Sum Rules and the lattice in  table \ref{tab:couplings}. As can be seen, there is reasonable agreement between the AdS/QCD prediction for the vector coupling $f_{\rho}$ and experiment. We note that our prediction for $f_{\rho}^{\perp}(\mu)$ hardly depends on $\mu$ for $\mu \ge 1$ GeV.  Viewed as a prediction  at some low scale $\mu \sim 1$ GeV, the agreement with Sum Rules and the lattice is reasonable. We note that the AdS/QCD prediction for the ratio of couplings is sensitive to the quark mass. For instance, using a current quark mass would yield a ratio far lower than the Sum Rules and lattice predictions. We therefore use here a constituent quark mass of $0.14$ GeV which is also the value used in \cite{Soyez:2007kg,Forshaw:2006np,Forshaw:2004vv}.

\begin{table}[h]
\begin{center}
\textbf{Couplings of the $\rho$ meson}
\[
\begin{array}
[c]{|c|c|c|c|c|c|}
\hline
\mbox{Reference}&\mbox{Approach} & \mbox{Scale}~\mu& f_{\rho} ~[\mbox{MeV}]&f_ {\rho}^{\perp} (\mu)~ [\mbox{MeV}] &f_\rho^{\perp}(\mu)/f_\rho \\ \hline
\mbox{\cite{Beringer:1900zz}}&\mbox{Experiment}& & 220 \pm 2 & & \\ \hline
\mbox{This paper}&\mbox{AdS/QCD} &\sim 1 ~ \mbox{GeV}&214&135&0.63 \\ \hline
\mbox{\cite{Ball:2006eu}}&\mbox{Sum Rules} &2~\mbox{GeV} &206 \pm 7 &145 \pm 8 & 0.70 \pm 0.04\\ \hline
\mbox{\cite{Becirevic:2003pn}}&\mbox{Lattice} &2 ~ \mbox{GeV} & &&0.72 \pm 0.02 \\ \hline
\mbox{\cite{Braun:2003jg}}&\mbox{Lattice} &2 ~ \mbox{GeV} & &&0.742 \pm 0.014 \\ \hline
\end{array}
\]
\end{center}
\caption {AdS/QCD predictions for the vector and tensor couplings of the $\rho$ meson compared to Sum Rules predictions, lattice predictions and experiment. We use $m_f=0.14$ GeV to make these predictions. }
\label{tab:couplings}
\end{table}

The twist-$2$ DAs can be expanded as \cite{Ball:1996tb,Ball:1998fj}
\begin{equation}
\phi_{\rho}^{||,\perp}(z,\mu)=6 z(1-z) \left [1 + \sum_{j=2,4,...} a_{j}^{||} (\mu)
C_{j}^{3/2}(\xi) \right] \;,
\label{gegenbauer_expansion}
\end{equation}
where $C_{j}^{3/2}(\xi)$ are the Gegenbauer polynomials and $\xi=2z-1$. Standard Sum Rules predictions are usually available only for $a_2^{\parallel,\perp}$. The twist-$2$ DAs are thus approximated as
\begin{equation}
\phi_{\rho}^{\parallel,\perp}(z,\mu) =  6 z(1-z) \left[ 1+
a_2^{\parallel,\perp} (\mu) \, \frac{3}{2} ( 5\xi^2  - 1 ) \right] 
\label{eq:phipar}
\end{equation}
i.e. by keeping only the first term in equation \eqref{gegenbauer_expansion}. We use here the Sum Rules estimates given in reference \cite{Ball:2007zt}: $a_2^{\parallel}=0.10\pm 0.05$ and $a_2^{\perp}=0.11\pm 0.05$ . Reference \cite{Ball:2007zt} also gives
explicit expressions for  the twist-$3$ DAs:
\begin{eqnarray}
  g^{\perp (v)}_{\rho}(z,\mu) & = & \frac{3}{4}(1+\xi^2)
 + \left(\frac{3}{7} \,
a_2^\parallel(\mu) + 5 \zeta_{3}(\mu) \right) \left(3\xi^2-1\right)
 \nonumber\\
& & {}+ \left[ \frac{9}{112}\, a_2^\parallel(\mu)
+ \frac{15}{32}\, \omega^{\parallel}_{3}(\mu) -\frac{15}{64}\tilde{\omega}^{\parallel}_{3}(\mu)
 \right] \left( 3 - 30 \xi^2 + 35\xi^4\right) \;.
\label{eq:gv}
\end{eqnarray}
and
\begin{equation}
g_{\rho}^{\perp (a)}(z,\mu)=6z(1-z)\left[ 1 + \left(\frac{1}{6} a_2^{\parallel}(\mu) + \frac{10}{9} \zeta_3^{\parallel}(\mu) + \frac{5}{12}\omega_3^{\parallel}(\mu) -\frac{5}{24} \tilde{\omega}_3^{\parallel}(\mu)\right) C_2^{3/2}(\xi) \right] \;.
\end{equation}
The Sum Rules estimates are $\zeta_3^{\parallel}(2~\mbox{GeV})=0.020\pm0.009$, $\omega^{\parallel}_{3}(2~\mbox{GeV}) =0.09\pm0.03$ and $\tilde{\omega}^{\parallel}_{3}(2~\mbox{GeV}) =-0.04\pm0.02$ \cite{Ball:2007zt}.

\begin{figure}
\centering
\subfigure[~Twist-$2$ DA for the longitudinally polarized $\rho$ meson]{\includegraphics[width=.60\textwidth]{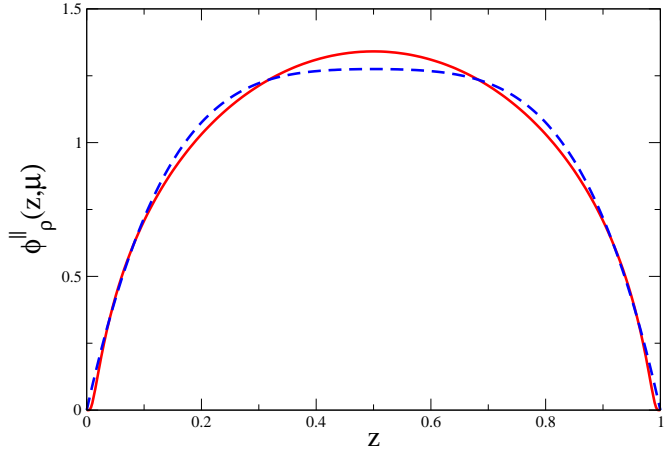} }
\subfigure[~Twist-$2$ DA for the transversely polarized $\rho$ meson]{\includegraphics[width=.60\textwidth]{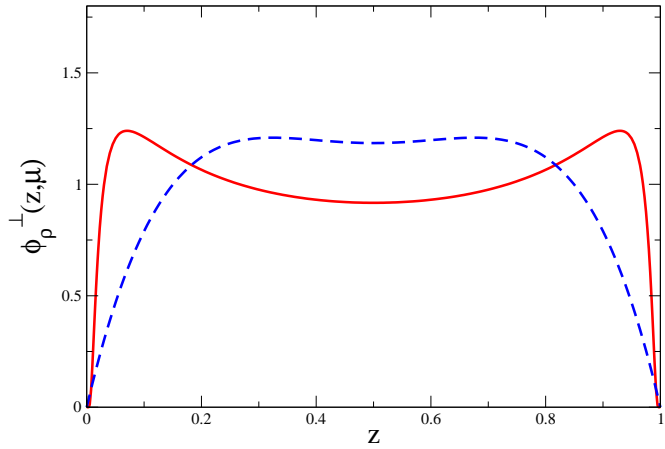} }
\caption{Twist-$2$ DAs for the $\rho$ meson. Solid Red: AdS/QCD DA at $\mu \sim 1$ GeV; Dashed Blue: Sum Rules DA at $\mu=2$ GeV.} \label{fig:tw2DAs}
\end{figure}

In figure \ref{fig:tw2DAs}, we compare the AdS/QCD twist-$2$  DAs to the SR twist-$2$ DAs at a scale $\mu=2$ GeV. We note that, as is the case for the AdS/QCD tensor coupling, the AdS/QCD DAs hardly depend on $\mu$ for $\mu \ge 1$ GeV and they should be viewed as parametrizations of the DAs at some low scale $\mu \sim 1$ GeV.  As can be seen and as was already noted in reference \cite{Forshaw:2012im}, the agreement between the AdS/QCD and Sum Rules twist-$2$ DA for the longitudinally polarized meson is good. On the other hand, we note different shapes for the AdS/QCD and Sum Rules predictions for the twist-$2$ DA of the transversely polarized meson. In particular, we find that the AdS/QCD DA has pronounced humps near the end-points and that when it starts decreasing, it does so faster than the SR DA. 

\begin{figure}
\centering
\subfigure[~Axial-vector twist-$3$ DA for the transversely polarized $\rho$ meson]{\includegraphics[width=.60\textwidth]{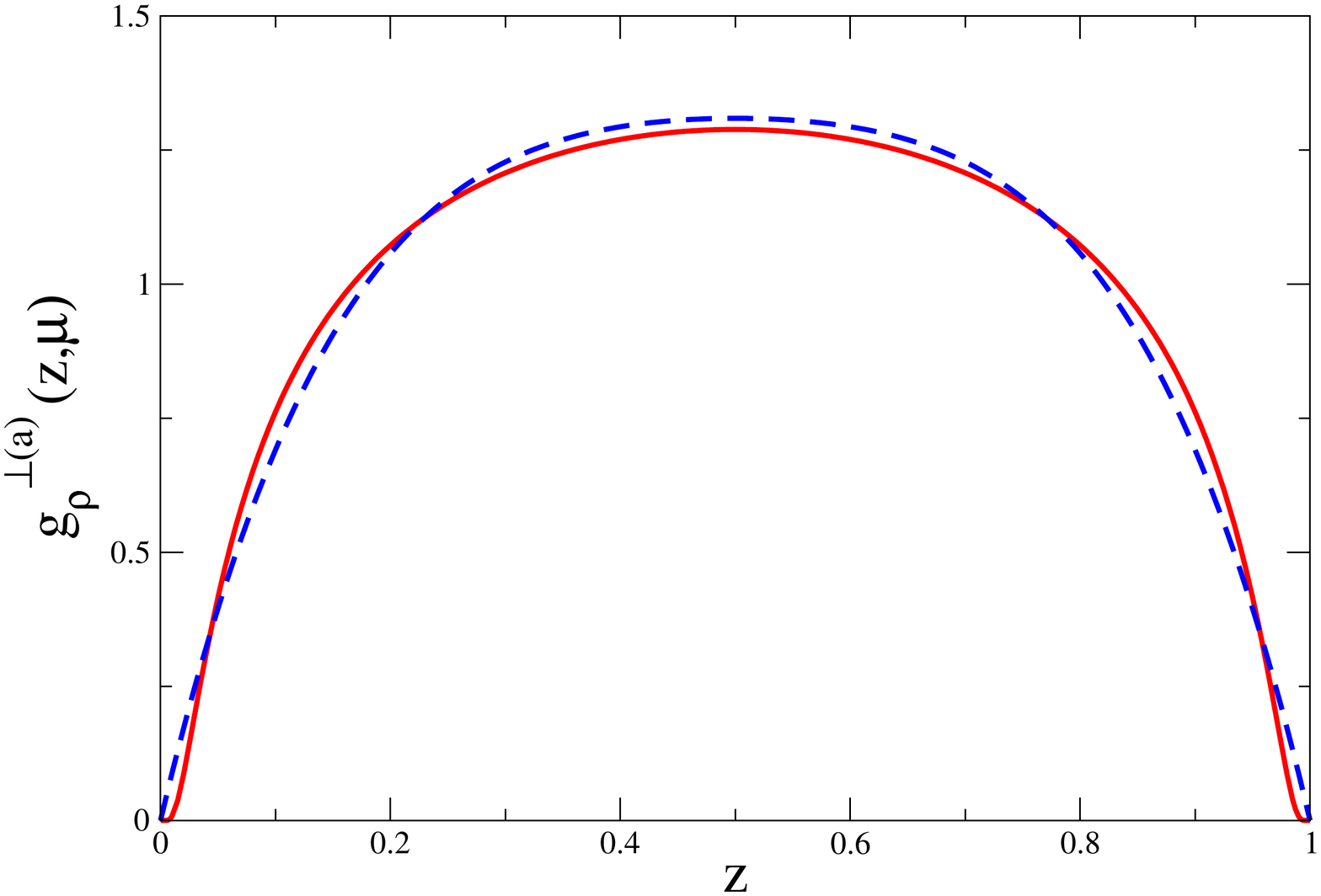} }
\subfigure[~Vector twist-$3$ DA for the transversely polarized $\rho$ meson]{\includegraphics[width=.60\textwidth]{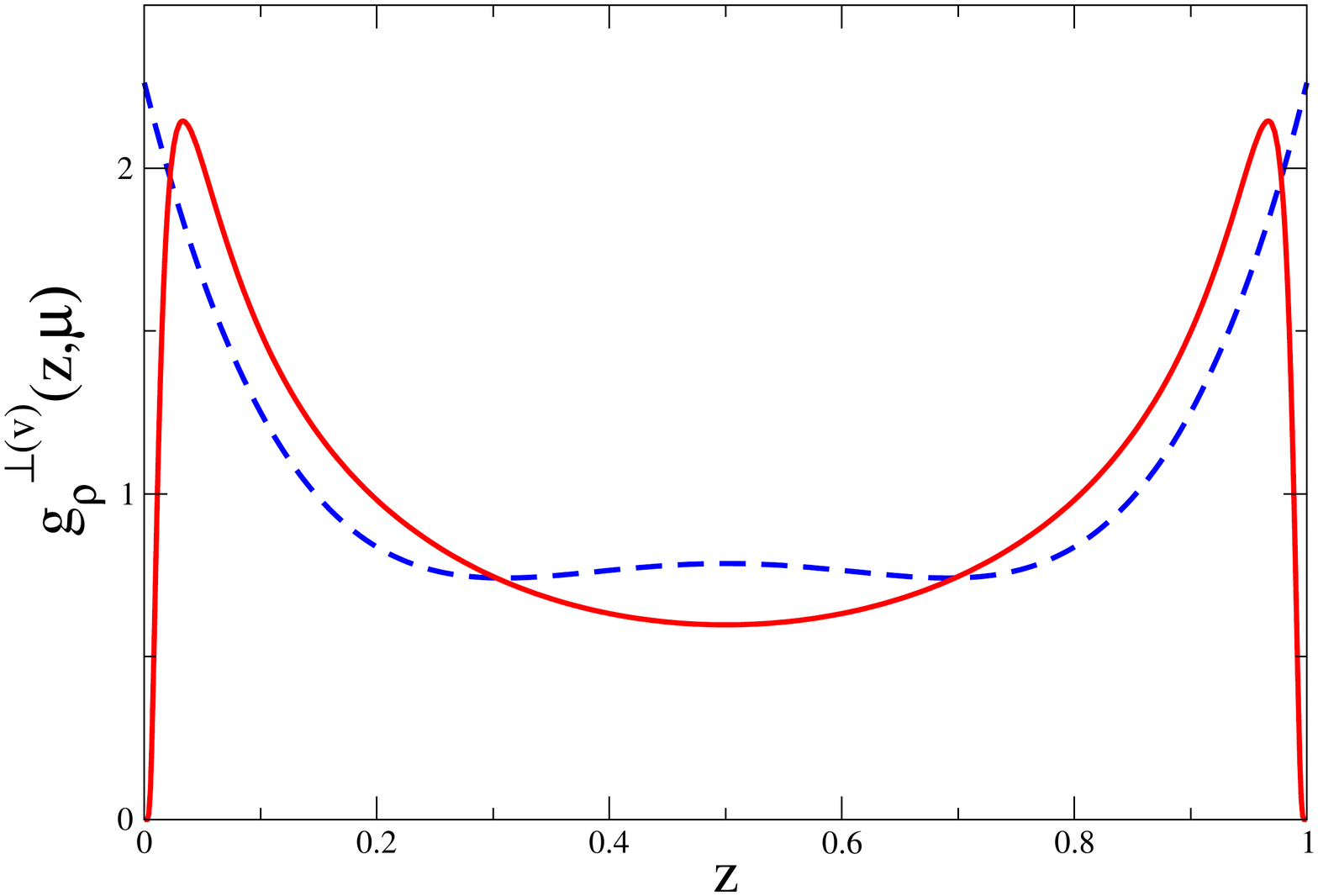} }
\caption{Twist-$3$ DAs for the $\rho$ meson. Solid Red: AdS/QCD DA at $\mu \sim 1$ GeV; Dashed Blue: Sum Rules DA at $\mu=2$ GeV.} \label{fig:tw3DAs}
\end{figure}

In figure \ref{fig:tw3DAs}, we compare AdS/QCD  twist-$3$  DAs to the SR twist-$3$ DAs at a scale $\mu=2$ GeV. The agreement between AdS/QCD and SR is quite good for both the axial vector DA but we note a difference between SR and AdS/QCD vector DA at the end-points: the AdS/QCD, unlike the  SR DA, falls to zero at the end-points.

\section{Branching ratios}
We are now in a position to compute the branching ratios given by 
\begin{equation}
\label{BR}
 {\cal BR}(\bar{B}^{\circ}_{(s)} \to \rho^\circ\gamma )=\frac{\tau_{B_{(s)}}}{16\pi M_{B_{(s)}}^3}\left (1-\left(\frac{M_\rho}{M_{B_{(s)}}}\right)^2\right )
|{\cal A} (\bar{B}^{\circ}_{(s)} \to\rho^\circ\gamma)|^2 \; ,
\end{equation}
where the amplitude ${\cal A} (\bar{B}^{\circ}_{(s)} \to\rho^\circ\gamma)$ is given by either equation \eqref{AtotB0} for $\bar{B}^{\circ} \to\rho^\circ\gamma$ or equation \eqref{AtotBs} for $\bar{B}^{\circ}_{s} \to\rho^\circ\gamma$ and 
$\tau_{B_{(s)}}$ is the measured lifetime of the $B_{(s)}$ meson \cite{Beringer:1900zz} . Before presenting our predictions for the branching ratios, it is instructive to compare the AdS/QCD and SR predictions for the  integrals $I_1^{\mbox{\tiny{tw2}}}(s_p,\mu)$ and  $I_2^{\mbox{\tiny{tw2}}}(\mu)$ given by equations \eqref{I1} and \eqref{I2} respectively. Our results are shown in table \ref{tab:I1I2}. 
We note that the integrals are not very sensitive to the precise shape of the twist-$2$ DA. 

\begin{table}[h]
\begin{center}
\[
\begin{array}
[c]{|c|c|c|}
\hline
\mbox{Integral}&\mbox{SR} &\mbox{AdS/QCD} \\ \hline
I_1^{\mbox{\tiny{tw2}}}(s_c,\mu)&1.902+2.620 i  & 1.590+2.329 i \\ \hline
I_1^{\mbox{\tiny{tw2}}}(s_u,\mu)& -6.561+0.030 i&-8.866+0.027i  \\ \hline
I_1^{\mbox{\tiny{tw2}}}(0,\mu)& -6.660  & -8.989 \\ \hline
I_2^{\mbox{\tiny{tw2}}} & 3.330& 4.495 \\ \hline
\end{array}
\]
\end{center}
\caption {AdS/QCD and SR predictions for the integrals $I_1^{\mbox{\tiny{tw2}}}$ and $I_2^{{\mbox{\tiny{tw2}}}}$ given by equations \eqref{I1} and \eqref{I2} respectively. The SR predictions are at a scale $\mu=2$ GeV and the AdS/QCD prediction are at a scale $\mu \sim 1$ GeV.}
\label{tab:I1I2}
\end{table}
We next compare the Sum Rules and the AdS/QCD predictions for the integrals $I_1^{\mbox{\tiny{tw3}}}$ and $I_2^{\mbox{\tiny{tw3}}}$  given by equations \eqref{I3} and \eqref{I4} respectively.  Our results are shown in table \ref{tab:I3I4B}. 
In this case, the AdS/QCD and SR predictions are drastically different.  The SR DA yields divergent integrals for both $I_1^{\mbox{\tiny{tw3}}}$ and $I_2^{\mbox{\tiny{tw3}}}$ unlike the AdS/QCD DA which leads to finite results in both cases. The divergent SR integrals could be estimated by introducing an IR cut-off but this procedure leads to a large uncertainty in the prediction for the annihilation amplitude \cite{Ahmady:2007ka}. 

\begin{table}[h]
\begin{center}
\[
\begin{array}
[c]{|c|c|c|}
\hline
 \mbox{Integral}&\mbox{SR} &\mbox{AdS/QCD} \\ \hline
I_{1(s)}^{\mbox{\tiny{tw3}}}(\mu)& \infty  &   0.237(0.229) \\ \hline
I_{2(s)}^{\mbox{\tiny{tw3}}}(\mu)& \infty & 0.036 (0.034)\\ \hline
\end{array}
\]
\end{center}
\caption {Sum Rules and AdS/QCD predictions for the integrals $I_1^{\mbox{\tiny{tw3}}}$ and $I_2^{{\mbox{\tiny{tw3}}}}$ given by equations \eqref{I3} and \eqref{I4} respectively. The SR predictions are at a scale $\mu=2$ GeV and the AdS/QCD prediction are at a scale $\mu \sim 1$ GeV.}
\label{tab:I3I4B}
\end{table}

\begin{figure}
\centering
\includegraphics[width=.60\textwidth]{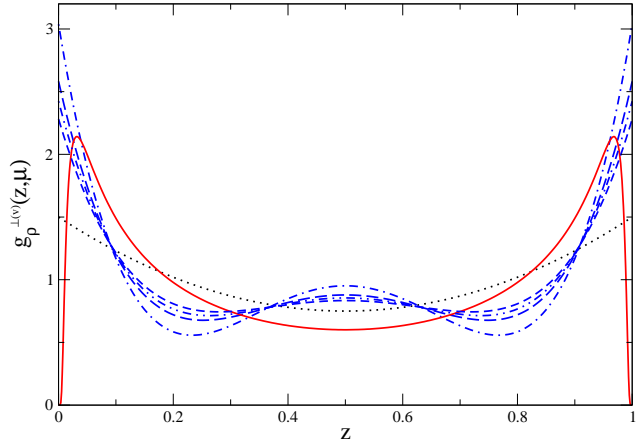} 
\caption{Evolution of the twist-$3$ SR DA. Blue: SR DA at $\mu=1$ (Dot-dashed), $2$ (Long-dashed), $3$ (Dot-dot-dashed) and $5$ (Short-dashed) GeV; Dotted Black: Asymptotic DA; Solid Red: AdS/QCD DA at $\mu \sim 1$ GeV.} \label{fig:SRevolve}
\end{figure}

It is instructive to investigate the influence of perturbative QCD scale evolution on the infrared divergence encountered with the SR DA. 
As shown in  figure \ref{fig:SRevolve}, we evolve the SR DA from $\mu=1$ GeV to $\mu=2, 3$ and $5$ GeV to leading logarithmic accuracy using the evolution in  \cite{Ball:1998ff}. We also show the asymptotic DA, i.e. the SR DA at $\mu \to \infty$.   As can be seen, the SR DAs do not vanish at the end-points and we find that the divergence problem persists at scales other than $2$ GeV. 
 Also shown in figure \ref{fig:SRevolve} is the AdS/QCD DA which, unlike the SR DA, vanishes at the end-points and avoids the end-point divergences. On the other hand,  the AdS/QCD DA lacks the perturbative evolution with the scale $\mu$ and must be viewed to be a parametrization of the DA at some low scale $\mu \sim 1$ GeV. This is a shortcoming of the AdS/QCD DA compared to the SR DA. However, we expect the AdS/QCD DA to be a reasonable parametrization of the DA at the scale $\mu=2$ GeV relevant to the decays we compute here although we cannot make a strong case that it will still avoid the end-point divergences if its perturbative QCD evolution with the scale $\mu$ is taken into account.

Our predictions for the branching ratio of $\bar{B}^{\circ} \to \rho^{\circ} \gamma$  are shown in table \ref{tab:BRB0}. In this table, we show how the predictions vary with the degree of accuracy of the calculation. The predicted branching ratio computed using the leading power amplitude at zeroth order in $\alpha_s$ (i.e. equation \eqref{LOA}) 
is clearly lower than the measured value. At this level of accuracy, the amplitude does not depend on the DAs.  The leading power amplitude becomes sensitive to the twist-$2$ DA at first order in $\alpha_s$ 
and at this level of accuracy, we find that the AdS/QCD and SR predictions agree with each other and with experiment. We confirm that all four power-suppressed annihilation contributions in $\bar{B}^{\circ} \to \rho^{\circ} \gamma$ are numerically small. Nevertheless, the AdS/QCD DA allows us to compute the annihilation contributions beyond leading power accuracy  without the ambiguity due to end-point divergences encountered with the SR DA. At the same time,  the AdS/QCD DA allows us to provide a more reliable theoretical estimate for  the branching ratio of the decay $\bar{B}_{s}^{\circ} \to \rho^{\circ} \gamma$ which proceeds mainly via annihilation and cannot be reliably predicted using the SR DA due to end-point divergences\cite{Ahmady:2007ka} . Using the AdS/QCD DAs, we predict a branching ratio of $5.5 \times 10^{-10}$ for this decay. This rare decay can be enhanced by NP  \cite{Ahmady:2007ka} and it would be interesting to investigate it at the LHCb.

\begin{table}[h]
\begin{center}
\textbf{Branching ratio ($\times 10^{-7}$) for $\bar{B}^\circ \to \rho^\circ \gamma$ }
\[
\begin{array}
[c]{|c|c|c|c|c|c|c|}
\hline
\mbox{DA}&\mbox{Accuracy}&\mbox{SR} &\mbox{AdS/QCD} & \mbox{PDG} & \mbox{Belle}  & \mbox{BaBar}\\ \hline
\mbox{tw}2 + \mbox{tw}3&\mbox{Lead.} (\alpha_s^1)+ \mbox{Anni.}[\alpha_s^0, (1/m_b)^2] &  &7.67&8.6 \pm 1.5  &7.8 \pm^{1.7}_{1.6} \pm^{0.9}_{1.0}& 9.7\pm^{2.4}_{2.2} \pm^{0.6}_{0.6}\\ \hline
\mbox{tw}2 &\mbox{Lead.}(\alpha_s^1) + \mbox{Anni.}[\alpha_s^0,(1/m_b)] & 7.86&7.65 & & & \\ \hline
\mbox{tw}2 &\mbox{Leading}~(\alpha^1_s) &7.87 &7.68 & & & \\ \hline
\mbox{None} &\mbox{Leading}~(\alpha_s^0)&4.76 & 4.76& & & \\ \hline
\end{array}
\]
\end{center}
\caption {Sum Rules and AdS/QCD predictions for the branching ratio ($\times 10^{-7}$) of $\bar{B}^\circ \to \rho^\circ \gamma$ using AdS/QCD or Sum Rules compared to the measurements from Belle \cite{Taniguchi:2008ty}, BaBar \cite{Aubert:2008al} and the average value from PDG \cite{Beringer:1900zz}.} 
\label{tab:BRB0}
\end{table}

\section{Conclusions}
We have used new holographic AdS/QCD DAs for the transversely polarised $\rho$ meson in order to compute the branching ratios for the decays $\bar B^\circ\rightarrow \rho^\circ \gamma$ and $B_s\rightarrow \rho^\circ \gamma$ beyond leading power accuracy. The AdS/QCD prediction for the branching ratio of $\bar B^\circ\rightarrow \rho^\circ \gamma$ agrees with experiment and we provide a theoretical estimate for the branching ratio of the rare decay $\bar B^\circ_s\rightarrow \rho^\circ \gamma$. We find that the AdS/QCD DAs are complementary to the standard SR DAs:  they agree with the SR predictions to leading power accuracy and  avoid  the end-point divergence ambiguity when computing some power corrections.  However, in its present form, the AdS/QCD DA lacks the perturbative QCD evolution and it remains to be seen if our conclusion remains valid if this evolution is taken into account.

\section{Acknowledgements}
This research is supported by the Natural Sciences and Engineering Research Council of Canada (NSERC). 

\bibliography{rho_revised4}

\end{document}